\documentclass[english,prb, aps, reprint]{revtex4-1}
\usepackage[T1]{fontenc}
\usepackage[latin9]{inputenc}
\usepackage{amstext}
\usepackage{amssymb}
\usepackage{graphicx}
\usepackage{esint}

\makeatletter

\newcommand{\lyxdot}{.}

\makeatother

\usepackage{babel}
\begin{document}
\title{Auger Recombination Lifetime Scaling for Type-I and Quasi-Type-II
Core/Shell Quantum Dots}
\author{John P. Philbin}
\email{jphilbin@berkeley.edu}

\affiliation{Department of Chemistry, University of California, Berkeley, California
94720, United States}
\author{Eran Rabani}
\email{eran.rabani@berkeley.edu}

\affiliation{Department of Chemistry, University of California, Berkeley, California
94720, United States}
\affiliation{Materials Sciences Division, Lawrence Berkeley National Laboratory,
Berkeley, California 94720, United States}
\affiliation{The Sackler Center for Computational Molecular and Materials Science,
Tel Aviv University, Tel Aviv, Israel 69978}
\begin{abstract}
Having already achieved near--unity quantum yields, with promising
properties for light--emitting diode, lasing, and charge separation
applications, colloidal core/shell quantum dots have great technological
potential. The shell thickness and band alignment of the shell and
core materials are known to influence the efficiency of these devices.
In many such applications, a key to improving the efficiency requires
a deep understanding of multiexcitonic states. Herein, we elucidate
the shell thickness and band alignment dependencies of the biexciton
Auger recombination lifetime for quasi--type--II CdSe/CdS and type--I
CdSe/ZnS core/shell quantum dots. We find that the biexciton Auger
recombination lifetime increases with the total nanocrystal volume
for quasi--type--II CdSe/CdS core/shell quantum dots and is independent
of the shell thickness for type--I CdSe/ZnS core/shell quantum dots.
In order to perform these calculations and compute Auger recombination
lifetimes, we developed a low--scaling approach based on the stochastic
resolution of identity. The numerical approach provided a framework
to study the scaling of the biexciton Auger recombination lifetimes
in terms of the shell thickness dependencies of the exciton radii,
Coulomb couplings, and density of final states in quasi--type--II
CdSe/CdS and type--I CdSe/ZnS core/shell quantum dots. 
\end{abstract}

\maketitle
The viability of many semiconductor nanomaterial--based applications
relies upon the ability to control multiexcitonic states.\citep{Klimov2000,Klimov2014}
For example, in typical nanomaterial--based lasers, generating population
inversion requires two excitons in the nanosystem and, thus, the properties
of the biexcitonic state determine, amongst other factors, the efficiency
of the device.\citep{Klimov2007,She2015,Lim2018,Roh2020} In fact,
this is arguably the case for other applications such as light--emitting
diodes\citep{Klimov2007,Bae2013b} and photocatalysts.\citep{Ben-Shahar2018}
Therefore, understanding the properties of the biexcitonic state and
its decay channels is central to improving and further developing
many light--induced applications.

\begin{figure*}
\includegraphics[width=16cm]{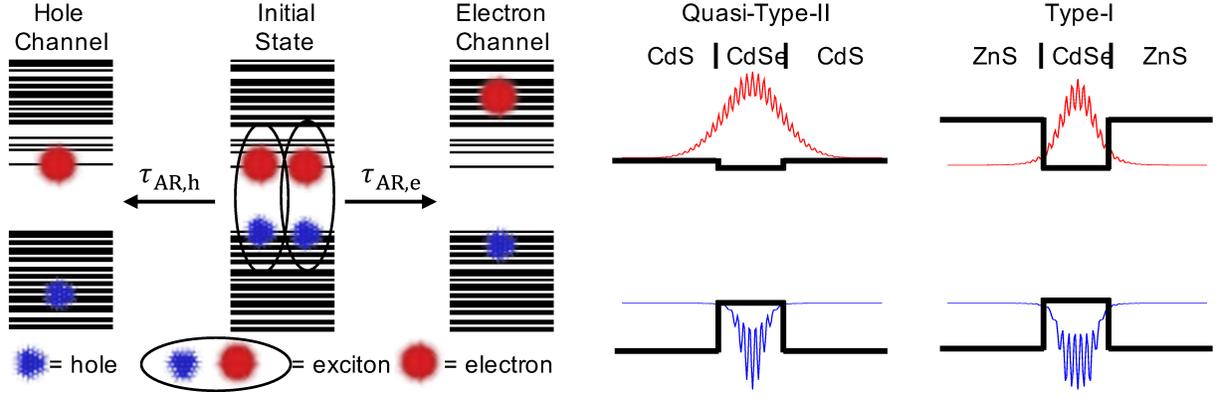}

\caption{(A) Schematic of an Auger recombination event. The initial biexcitonic
state is shown as two spatially uncorrelated excitonic states and
the final states are shown as unbound electron--hole pairs. The hole
(electron) channel on the left (right) shows the hole (electron) receiving
a majority of the energy from the recombining exciton. (b) Schematic
of the quasi--type--II nature of CdSe/CdS core/shell quantum dots
and the type--I nature of CdSe/ZnS core/shell quantum dots. Projected
electron (red) and hole (blue) probability densities are shown on
top of the band alignment scheme to highlight the differences in electron
localization between the two systems.\label{fig:band_alignment_AR_scheme}}
\end{figure*}

One of the major decay channels of the biexcitonic state is Auger
recombination (similar to exciton--exciton annihilation), which is
a nonradiative process where an electron and hole recombine and transfer
their energy to a nearby electron or hole in a Coulomb mediated process
(Figure\ \ref{fig:band_alignment_AR_scheme}). Auger recombination
is typically is the dominant decay channel of biexcitons in semiconductor
nanocrystals as it usually occurs on a sub--nanosecond timescale.

An aspect of biexciton Auger recombination that has drawn much attention
over the years is that of how the rate of biexciton Auger recombination
decay depends on the size of the nanocrystal.\citep{Klimov2000,Klimov2008,Robel2009,Padilha2013,Vaxenburg2015,Li2017,Philbin2018,Li2019}
For single material colloidal quantum dots (QDs), the linear dependence
of the biexciton lifetime with the QD volume has become known as the
``universal volume scaling law.''\citep{Robel2009} Although the
size of a single material colloidal QD is a knob that can be tuned
to change the biexciton lifetime and, thus, the efficiency of nanodevices
that rely on biexcitonic states, changing the size also drastically
impacts single exciton properties. On the other hand, heterostructure
nanomaterials have many experimentally tunable parameters, including
relative size and band alignments between the individual component
materials, that can be chosen to optimize the performance of nanodevices.
For example, independently tuning the shell thickness and band alignment
has resulted in heterostructure nanocrystals with near--unity quantum
yields along with promising light--emitting diode and lasing properties.\citep{Klimov2007,Garcia-Santamaria2009,Bae2013b,Chen2013,Hadar2017,Hazarika2019,Hanifi2019}
Interestingly, there have been multiple reports that the ``universal
volume scaling law'' does not apply to core/shell QDs.\citep{Garcia-Santamaria2011,Bae2013b,Javaux2013,Vaxenburg2016,Pelton2017,Zhang2018,Kong2018}
Although, significant theoretical progress has been made,\citep{Jain2016,Kaledin2018}
particularly on the impact of the sharpness of the core/shell interface
on biexciton lifetimes,\citep{Cragg2010} a quantitatively accurate
atomistic electronic structure method has not yet been developed for
heterostructure nanomaterials due to the inherently large nature of
heterostructure nanosystems and the steep scaling with system size
of computing Auger recombination lifetimes. 

With this difficulty in mind, we report an efficient, stochastic method
for calculating biexciton Auger recombination (AR) lifetimes within
Fermi's golden rule suitable for large heterostructure nanosystems
and apply it to elucidate the shell thickness dependence of AR in
quasi--type--II CdSe/CdS and type--I CdSe/ZnS core/shell QDs (Figure\ \ref{fig:band_alignment_AR_scheme}).
The stochastic approach, which also accounts for electron--hole correlations,
reduces the scaling with the system size ($N$) of calculating AR
lifetimes from $O\left(N^{5}\right)$ to $O\left(N^{2}\right)$ and
predicts quantitatively accurate AR lifetimes in comparison to experiments.
Additionally, the AR formalism predicts that adding a shell with a
quasi--type--II band alignment (CdSe/CdS QDs) results in an increase
in the AR lifetime, in agreement with previous experimental and theoretical
results,\citep{Park2014,Vaxenburg2016,Zhang2018,Kong2018,Kaledin2018}
whereas the addition of a shell with a strictly type--I band alignment
(CdSe/ZnS QDs) has little impact on the AR lifetime. Lastly, we explain
the shell thickness dependencies of the AR lifetimes in terms of the
size dependencies of the root--mean--square exciton radius, Coulomb
coupling, and density of final states in quasi--type--II CdSe/CdS
and type--I CdSe/ZnS core/shell QDs.

AR is a Coulomb mediated process for which an initial biexcitonic
state ($\left|B\right\rangle $) of energy $E_{B}$ decays into a
final excitonic state ($\left|S\right\rangle $) of energy $E_{S}$
via Coulomb scattering ($V$). An AR lifetime ($\tau_{\text{AR}}$)
for a nanomaterial can be calculated using Fermi's golden rule where
we average over thermally distributed initial biexcitonic states and
sum over all final decay channels into single excitonic states:
\begin{eqnarray}
\tau_{\text{AR}}^{-1} & = & \sum_{B}\frac{e^{-\beta E_{B}}}{Z_{B}}\left[\frac{2\pi}{\hbar}\sum_{S}\left|\left\langle B\left|V\right|S\right\rangle \right|^{2}\delta\left(E_{B}-E_{S}\right)\right].\label{eq:fermisGoldenRule}
\end{eqnarray}
In the above, the delta function ($\delta\left(E_{B}-E_{S}\right)$)
enforces energy conservation between the initial and final states
and the partition function ($Z_{B}=\sum_{B}e^{-\beta E_{B}}$) is
for the initial biexcitonic states (we assume Boltzmann statistics
for biexcitons). Utilizing the interacting framework, previously developed
by Philbin and Rabani,\citep{Philbin2018} a deterministic calculation
of an AR lifetime can be performed using
\begin{widetext}
\begin{eqnarray}
\tau_{\text{AR}}^{-1} & = & \frac{2\pi}{\hbar Z_{B}}\sum_{B}e^{-\beta E_{B}}\sum_{a,i}\left|\sum_{b,c,k}c_{b,i}^{B}c_{c,k}^{B}V_{abck}\right|^{2}\delta\left(E_{B}-\varepsilon_{a}+\varepsilon_{i}\right)\label{eq:dIntAR}\\
 &  & +\frac{2\pi}{\hbar Z_{B}}\sum_{B}e^{-\beta E_{B}}\sum_{a,i}\left|\sum_{j,c,k}c_{a,j}^{B}c_{c,k}^{B}V_{ijck}\right|^{2}\delta\left(E_{B}-\varepsilon_{a}+\varepsilon_{i}\right),\nonumber 
\end{eqnarray}
\end{widetext}

\noindent where the indices $a,b,c...$ refer to the electron (unoccupied)
states, $i,j,k...$ refer to the hole (occupied) states with corresponding
energies $\varepsilon_{a}$ and $\varepsilon_{i}$, $r,s,u...$ are
general indices, and $V_{rsut}$ is the Coulomb coupling given by
\begin{eqnarray}
V_{rsut} & = & \iint\frac{\phi_{r}\left(\mathbf{r}\right)\phi_{s}\left(\mathbf{r}\right)\phi_{u}\left(\mathbf{r^{\prime}}\right)\phi_{t}\left(\mathbf{r}^{\prime}\right)}{\left|\mathbf{r}-\mathbf{r}^{\prime}\right|}d^{3}\mathbf{r}\,d^{3}\mathbf{r}^{\prime}.\label{eq:coulombCoupling}
\end{eqnarray}
The coefficients ($c_{c,k}^{B}$) in Eq.\ (\ref{eq:dIntAR}) are
determined by solving the Bethe--Salpeter equation.\citep{Rohlfing2000}
For more details, please consult Ref.~\citenum{Philbin2018}. The
above formalism includes spatial correlations within the electron--hole
pairs but ignores them between the two excitons~\citep{Refaely-Abramson2017}
and in the final electron--hole pair (Figure\ \ref{fig:band_alignment_AR_scheme}).
This approximation for the final state is valid in a majority of nanomaterials
as the energy of the final electron--hole pair is approximately twice
the optical gap, which is well above the typical exciton binding energy
in all semiconductor nanomaterials.\citep{Wang2006} In other words,
the criteria for being able to approximate the final high energy excitonic
state as an uncorrelated electron--hole pair instead of a Wannier
or Frenkel exciton is that $E_{\text{opt}}\gg E_{\text{b}}$, where
$E_{\text{opt}}$ is the optical gap and $E_{\text{b}}$ is the exciton
binding energy. It was previously shown that this interacting (i.e.
exciton--based) AR formalism (Eq.\ (\ref{eq:dIntAR})) predicts
quantitively accurate AR lifetimes for both single material QDs and
nanorods.\citep{Philbin2018} On the other hand, noninteracting formalisms
that ignore all electron--hole correlations in the initial biexcitonic
state predict neither accurate AR lifetimes nor the scaling of the
lifetimes with respect to QD volume except for QDs in the very strong
confinement regime~\citep{Wang2003} --- highlighting the importance
of electron--hole correlations and the resulting Wannier exciton
formation in semiconductor nanomaterials.\citep{Philbin2018} 

The main drawback of the exciton--based (interacting) AR formalism
for calculating AR lifetimes (Eq.\ (\ref{eq:dIntAR})) is the computational
cost. Formally, the steepest scaling involved in Eq.\ (\ref{eq:dIntAR})
is diagonalization of the Bethe--Salpeter Hamiltonian to obtain the
coefficients ($c_{c,k}^{B}$), which formally scales as $O\left(N^{6}\right)$.
However, in practice this takes less than $10\%$ of the computational
time for nanomaterials with $\leq10,000$ atoms because only a few
low--lying energy states are required in order to calculate the AR
lifetime due to the Boltzmann factors in Eq.\ (\ref{eq:dIntAR}).
The majority of the computational time is spent on calculating all
of the Coulomb matrix elements, $V_{abck}$ and $V_{ijck}$, that
couple the initial biexcitonic states with the final electron--hole
pairs. The number of Coulomb matrix elements that must be calculated
scales as $O\left(N_{\text{e,final}}N_{\text{e}}^{2}N_{\text{h}}+N_{\text{h,final}}N_{\text{h}}^{2}N_{\text{e}}\right)\sim O\left(N^{4}\right)$,
where $N_{\text{e(h),final}}$ is the number of high energy final
electron (hole) states and $N_{\text{e(h)}}$ is the number of band--edge
electron (hole) states, and the cost of calculating each Coulomb matrix
element scales with the number of real--space grid points ($N_{\text{grid}}$)
as $O\left(N_{\text{grid}}\ln N_{\text{grid}}\right)$ to give the
overall scaling of $O\left(N^{5}\right)$. This limits the application
of Eq.\ (\ref{eq:dIntAR}) to relatively small systems ($<1,000$
atoms). 

To reduce the computational effort and scaling of the rate limiting
step, we employ a plane--wave stochastic representation of the Coulomb
operator:\citep{Neuhauser2016} 
\begin{eqnarray}
V_{rsut} & \approx & \left\langle R_{rs}^{\zeta}R_{ut}^{\zeta}\right\rangle _{\zeta}\label{eq:stoch_coulomb_rep}
\end{eqnarray}
where the notation $\left\langle ...\right\rangle _{\zeta}$ denotes
an average over $N_{\text{s}}$ stochastic orbitals (defined below),

\begin{eqnarray}
R_{rs}^{\zeta} & = & \int\phi_{r}^{*}\left(\mathbf{r}\right)\phi_{s}^{*}\left(\mathbf{r}\right)\theta^{\zeta}\left(\mathbf{r}\right)\,d^{3}\mathbf{r},
\end{eqnarray}
and $\theta^{\zeta}\left(\mathbf{r}\right)$ is a stochastic representation
of the Coulomb integral given by

\begin{eqnarray}
\theta^{\zeta}\left(\mathbf{r}\right) & = & \frac{1}{\left(2\pi\right)^{3}}\int d\mathbf{k}\,\sqrt{\widetilde{u}_{\text{C}}\left(\mathbf{k}\right)}\,\text{e}^{\text{i}\varphi\left(\mathbf{k}\right)}\,\text{e}^{\text{i}\mathbf{k\cdot r}}.
\end{eqnarray}
In the above equations, $\varphi\left(\mathbf{k}\right)$ is a random
phase between $0$ and $2\pi$ at each $k$--space grid point, $\widetilde{u}_{\text{C}}\left(\mathbf{k}\right)=\frac{4\pi}{k^{2}}$
is the Fourier transform of the Coulomb potential, and the stochastic
orbitals ($\theta^{\zeta}\left(\mathbf{r}\right)$) are indexed by
$\zeta$. By inserting Eq.\ (\ref{eq:stoch_coulomb_rep}) into Eq.\ (\ref{eq:dIntAR}),
we obtain 
\begin{widetext}
\begin{eqnarray}
\tau_{\text{AR}}^{-1} & = & \tau_{\text{AR,e}}^{-1}+\tau_{\text{AR,h}}^{-1}\label{eq:sARCoulomb}\\
\tau_{\text{AR,e}}^{-1} & = & \frac{2\pi}{\hbar Z_{B}}\sum_{B}e^{-\beta E_{B}}\sum_{a,i}\left\langle \sum_{b}c_{b,i}^{B}R_{ab}^{\zeta^{\prime}}\sum_{c,k}c_{c,k}^{B}R_{ck}^{\zeta^{\prime}}\right\rangle _{\zeta^{\prime}}^{*}\left\langle \sum_{b}c_{b,i}^{B}R_{ab}^{\zeta}\sum_{c,k}c_{c,k}^{B}R_{ck}^{\zeta}\right\rangle _{\zeta}\delta\left(E_{B}-\varepsilon_{a}+\varepsilon_{i}\right)\nonumber \\
\tau_{\text{AR,h}}^{-1} & = & \frac{2\pi}{\hbar Z_{B}}\sum_{B}e^{-\beta E_{B}}\sum_{a,i}\left\langle \sum_{j}c_{a,j}^{B}R_{ij}^{\zeta^{\prime}}\sum_{c,k}c_{c,k}^{B}R_{ck}^{\zeta^{\prime}}\right\rangle _{\zeta^{\prime}}^{*}\left\langle \sum_{j}c_{a,j}^{B}R_{ij}^{\zeta}\sum_{c,k}c_{c,k}^{B}R_{ck}^{\zeta}\right\rangle _{\zeta}\delta\left(E_{B}-\varepsilon_{a}+\varepsilon_{i}\right),\nonumber 
\end{eqnarray}
\end{widetext}

\noindent where $\tau_{\text{AR,e}}$ and $\tau_{\text{AR,h}}$ are
the lifetimes for the electron and hole channels, respectively (Figure\ \ref{fig:band_alignment_AR_scheme}).
The calculation of an AR lifetime using Eq.\ (\ref{eq:sARCoulomb})
scales as $O\left(N^{3}\right)$. 

To further reduce the computational scaling and complexity, we utilize
the stochastic resolution of the identity~\citep{Takeshita2017,Dou2019}
within the subspace of the final high energy electron and hole parts
of the Hamiltonian. In simpler terms, we sample the final high energy
electron and hole states in order to reduce the scaling with number
of final excitonic states. Thus, we arrive at a general expression
for calculating AR lifetimes of semiconductor nanomaterials using
an efficient, doubly stochastic formulation of the interacting (exciton--based)
AR formalism
\begin{widetext}
\begin{eqnarray}
\tau_{\text{AR}}^{-1} & = & \tau_{\text{AR,e}}^{-1}+\tau_{\text{AR,h}}^{-1}\label{eq:dsAR}\\
\tau_{\text{AR,e}}^{-1} & = & \frac{2\pi}{\hbar Z_{B}}\sum_{B}e^{-\beta E_{B}}\left\langle \left\langle \sum_{b}c_{b,i^{A}}^{B}R_{\theta^{A}b}^{\zeta^{\prime}}\sum_{c,k}c_{c,k}^{B}R_{ck}^{\zeta^{\prime}}\right\rangle _{\zeta^{\prime}}^{*}\left\langle \sum_{b}c_{b,i^{A}}^{B}R_{\theta^{A}b}^{\zeta}\sum_{c,k}c_{c,k}^{B}R_{ck}^{\zeta}\right\rangle _{\zeta}\right\rangle _{A}\nonumber \\
\tau_{\text{AR,h}}^{-1} & = & \frac{2\pi}{\hbar Z_{B}}\sum_{B}e^{-\beta E_{B}}\left\langle \left\langle \sum_{j}c_{a^{I},j}^{B}R_{\theta^{I}j}^{\zeta^{\prime}}\sum_{c,k}c_{c,k}^{B}R_{ck}^{\zeta^{\prime}}\right\rangle _{\zeta^{\prime}}^{*}\left\langle \sum_{j}c_{a^{I},j}^{B}R_{\theta^{I}j}^{\zeta}\sum_{c,k}c_{c,k}^{B}R_{ck}^{\zeta}\right\rangle _{\zeta}\right\rangle _{I},\nonumber 
\end{eqnarray}
\end{widetext}

\noindent where the indices $\theta^{A},i^{A}$ and $a^{I},\theta^{I}$
in Eq.\ (\ref{eq:dsAR}) are sampled final states from the complete
set of single excitonic states ($a,i$ pairs) in Eq.\ (\ref{eq:sARCoulomb}).
Energy conservation in Eq.\ (\ref{eq:dsAR}) has been taken into
account when forming the stochastic orbitals that sample the final
excitonic states, namely, we only sample states that preserve energy.
The computational cost of Eq.\ (\ref{eq:dsAR}) is $O\left(N^{2}\right)$.
This scaling does assume that the number of stochastic orbitals required
to properly converge the calculations does not increase with the system
size, which has shown to be true for a variety of electronic structure
methods.\citep{Baer2012,Baer2013a,Neuhauser2014b,Takeshita2017,Dou2019}
Another beneficial feature of Eq.\ (\ref{eq:dsAR}) is that it is
embarrassingly parallel over all sets of stochastic orbitals. The
speedup that arises from using Eq.\ (\ref{eq:dsAR}) instead of Eq.\ (\ref{eq:dIntAR})
ranges from $\sim5$ for QDs with $1,000$ atoms to greater than $1,000$
for QDs with $10,000$ atoms. This speedup made the study of the large
core/shell QDs presented in the remainder of this Letter possible.
The Supporting Information contains more information on the derivation,
implementation and computational cost of the above equations and similar
expressions for the noninteracting, free carrier--based formalism. 

We have implemented the above equations using the semi--empirical
pseudopotential method to model the electron and hole states.\citep{Wang1994,Wang1996,Rabani1999b,Wang2003}
We utilized the filter--diagonalization technique~\citep{Wall1995,Toledo2002}
to selectively calculate the low energy electron and hole states required
to accurately describe the excitonic states that compose the initial
biexcitonic state and the high energy electron and hole states that
satisfy energy conservation. The Bethe--Salpeter equation~\citep{Rohlfing2000}
was solved within the static screening approximation. And all electronic
structure calculations were performed using the minimum energy atomic
configuration obtained via molecular dynamic minimization~\citep{Zhou2013}
of the heterostructure QDs. This computational scheme has been shown
to predict quantitatively accurate single excitonic properties (e.g.
optical gap and emission polarizations) and accurately takes into
account the important effects of strain in heterostructure nanomaterials
that arise from the lattice mismatch between core and shell materials.\citep{Hadar2017,Hazarika2019}

\begin{figure}
\includegraphics[angle=270,width=8cm]{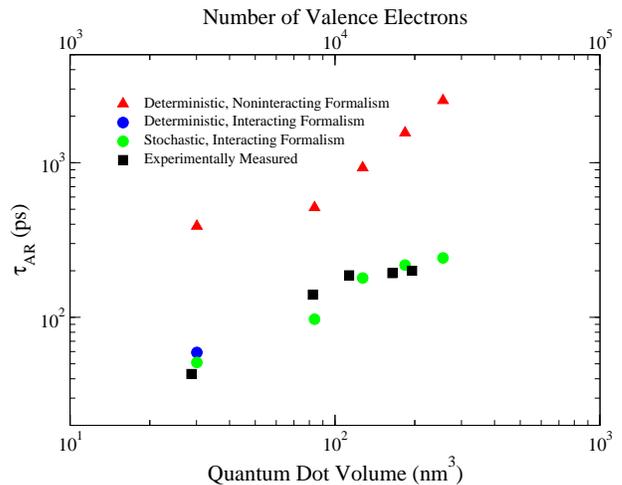}\caption{Comparison of Auger recombination lifetimes ($\tau_{\text{AR}}$)
that have been experimentally measured~\citep{Kong2018} (black),
calculated using the deterministic interacting formalism (blue), stochastic
formulation of the interacting formalism (green), and the deterministic
formulation of the noninteracting formalism (red) of CdSe/CdS core/shell
quantum dots with a CdSe core diameter of $3.8\:\text{nm}$ and varying
number of CdS shell monolayers.\label{fig:d3.8ARLifetimes}}
\end{figure}

Figure\ \ref{fig:d3.8ARLifetimes} displays the calculated AR lifetimes
using Eq.\ (\ref{eq:dsAR}) for the $d_{\text{core}}=3.8\text{ nm}$
CdSe/CdS QDs along with the experimentally measured AR lifetimes~\citep{Kong2018}
and AR lifetimes calculated using a noninteracting, free carrier--based
formalism.\citep{Wang2003} Quantitative agreement with the experimental
measurements on similarly sized CdSe/CdS QDs is observed when Eq.\ (\ref{eq:dsAR})
is used. It is important to note that all of the core/shell QDs studied
in this work have sharp core/shell interfaces.\citep{Zhang2018} In
other words, there is no alloying region between the core and shell
materials that is known to have important consequences on AR lifetimes.\citep{Cragg2010,Garcia-Santamaria2011,Jain2016}
The quantitative agreement shows the generality of the interacting
(exciton--based) AR formalism for predicting quantitatively accurate
AR lifetimes in nanomaterials. It is worthwhile to note that a noninteracting
(free--carrier based) AR formalism predicts incorrect AR lifetimes
in core/shell QDs, similar to the single material case.

\begin{figure}
\includegraphics[angle=270,width=8cm]{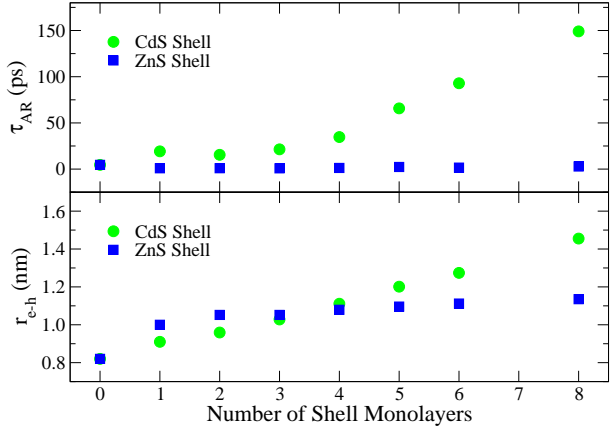}\caption{Auger recombination lifetimes (top) and root--mean--square exciton
radii ($\text{r}_{\text{e-h}}=\sqrt{\left\langle r_{\text{e-h}}^{2}\right\rangle }$),
bottom) of CdSe/CdS (green) and CdSe/ZnS (blue) core/shell quantum
dots as a function of the number CdS and ZnS shell monolayers, respectively,
for a CdSe core diameter of $2.2\text{ nm}$.\label{fig:ar_lifetimes_cds_zns}}
\end{figure}

Figure\ \ref{fig:ar_lifetimes_cds_zns} summarizes a main result
of this work. The top panel of Figure\ \ref{fig:ar_lifetimes_cds_zns}
compares calculated AR lifetimes for CdSe cores with a diameter of
$2.2\text{ nm}$ ($d_{\text{core}}=2.2\text{ nm}$) as a function
of the number of shell monolayers (MLs) for both CdS and ZnS from
$0\:\text{MLs}$ to up to $8\text{ MLs}$. This constitutes a range
of nanocrystal sizes from approximately $200$ atoms ($V_{\text{QD}}\sim5\text{ nm}^{3}$)
to nearly $10,000$ atoms ($V_{\text{QD}}\sim350\text{ nm}^{3}$).
Figure\ \ref{fig:ar_lifetimes_cds_zns} highlights the dramatically
different impact that growing a quasi--type--II shell (CdS) has
on the AR lifetime compared to growing a type--I (ZnS) shell on a
QD core (CdSe). Specifically, the addition of more and more CdS MLs
leads to the AR lifetime increasing from $\sim5\text{ ps}$ for the
$0\text{ ML}$ QD to $\sim35\text{ ps}$ and $\sim150\text{ ps}$
upon addition of $4$ and $8\text{ MLs}$ of CdS, respectively, for
the $d_{\text{core}}=2.2\text{ nm}$ CdSe QD core. On the other hand,
for the same CdSe core, the addition of $4$ and $8\text{ MLs}$ of
ZnS does not lead to an increase in the AR lifetime. 

In order to understand the vastly different shell thickness dependencies
of the AR lifetimes between CdSe/CdS and CdSe/ZnS QDs, we plot the
root--mean--square exciton radius ($\sqrt{\left\langle r_{\text{e-h}}^{2}\right\rangle }$
where $r_{\text{e-h}}$ is the electron--hole radial coordinate)~\citep{Brumberg2019}
as a function of the number of shell MLs for both series of core/shell
QDs in the bottom panel of Figure\ \ref{fig:ar_lifetimes_cds_zns}.
For CdSe/CdS QDs, the root--mean--square exciton radius systematically
increases with the number of shell MLs. On the other hand, for CdSe/ZnS
QDs there is an increase upon adding the first ZnS layer, but then
the addition of more and more ZnS MLs barely changes the root--mean--square
exciton radius. Specifically, the root--mean--square exciton radius
increases from $1.11\text{ nm}$ to $1.46\text{ nm}$ upon going from
$4\text{ MLs}$ to $8\text{ MLs}$ of CdS but only increases from
$1.08\text{ nm}$ to $1.14\text{ nm}$ upon going from $4\text{ MLs}$
to $8\text{ MLs}$ of ZnS for the same $d_{\text{core}}=2.2\text{ nm}$
CdSe core (bottom panel of Figure\ \ref{fig:ar_lifetimes_cds_zns}). 

\begin{figure}
\includegraphics[angle=270,width=8cm]{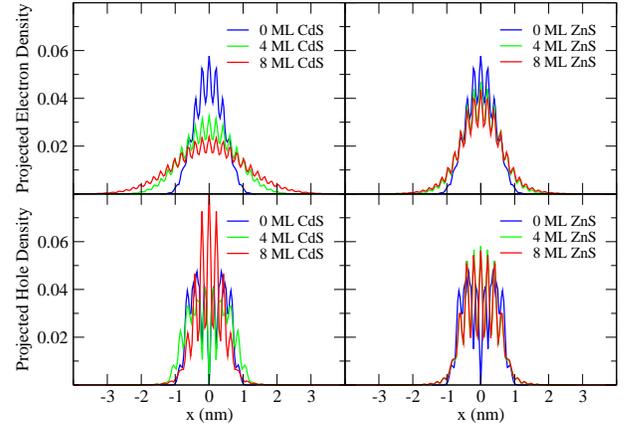}\caption{Hole and electron carrier densities of the lowest energy excitonic
state for a series of shell thicknesses for CdSe/CdS and CdSe/ZnS
core/shell QDs with a CdSe core diameter of $2.2\text{ nm}$.\label{fig:hole_electron_densities}}
\end{figure}

These different dependencies of the AR lifetime and root--mean--square
exciton radius with shell thickness are a direct consequence of the
quasi--type--II~\citep{Eshet2013,Kong2018} and type--I nature
of the CdS and ZnS shells, respectively. Figure\ \ref{fig:hole_electron_densities}
shows the hole and electron carrier densities of the lowest energy
excitonic state (i.e. electron--hole interactions have been included)
projected onto the x--axis of the core/shell QDs for the $d_{\text{core}}=2.2\text{ nm}$
CdSe QD cores with $0\text{ MLs}$, $4\text{ MLs}$ and $8\text{ MLs}$
of shell. For CdSe/CdS (left panels of Figure\ \ref{fig:hole_electron_densities}),
the quasi--type--II nature can be observed as the projected hole
density remains confined to the CdSe core for all shell thicknesses
while the electron density continuously spreads out into the CdS shell.
In contrast, both the hole and electron densities remain confined
to the CdSe core in CdSe/ZnS core/shell QDs, highlighting the type--I
band alignment of CdSe/ZnS core/shell QDs (right panels of Figure\ \ref{fig:hole_electron_densities}).
The impact of the electron spreading out into the CdS shell and, thus,
increasing the root--mean--square exciton radius in larger CdS shell
nanocrystals leads to a decrease in the Coulomb coupling involved
in AR calculations. This result can be understood by noting that the
larger the electron and hole wavefunctions overlap the larger the
Coulomb matrix elements, as the product $\phi_{c}\left(\mathbf{r^{\prime}}\right)\phi_{k}\left(\mathbf{r}^{\prime}\right)$
where $\phi_{c}\left(\mathbf{r^{\prime}}\right)$ and $\phi_{k}\left(\mathbf{r}^{\prime}\right)$
are wavefunctions for an initial electron and hole, respectively,
arises in the Coulomb coupling (Eq.\ (\ref{eq:coulombCoupling})).

The type--I band alignment of CdSe/ZnS core/shell QDs results in
the addition of ZnS MLs barely changing the root--mean--square exciton
radius and not increasing the AR lifetime. Surprisingly, the AR lifetimes
for all CdSe/ZnS core/shell QDs are slightly shorter, with lifetimes
of $\sim2\text{ ps}$, compared to the $\sim5\text{ ps}$ AR lifetime
for the bare CdSe core (Figure\ \ref{fig:ar_lifetimes_cds_zns}).
To elucidate whether or not the compressive strain of the ZnS shell
causes the decrease of the AR lifetime, we performed AR lifetimes
calculations on strained CdSe cores. Specifically, we performed molecular
dynamics based structural minimizations with ZnS shells and then removed
the ZnS shells before performing the electronic structure calculations.
This procedure resulted in compressively strained CdSe QDs,\citep{Hazarika2019}
where the degree of compressive strain was related to the number of
ZnS MLs that were present during molecular dynamics minimization.
Our calculations on this series of CdSe QDs show that the AR lifetime
decreases from $\sim5\text{ ps}$ to $\sim2\text{ ps}$ upon increasing
the strain on the CdSe QD (Table SX). Interestingly, the AR lifetime
decreasing by $\sim250\%$ upon adding strain to the CdSe QD is much
greater than would be expected due to just a volumetric change as
the compressive strain only changes the CdSe QD volume by $\sim10\%$.
We were able to trace the decrease of the AR lifetime to a decrease
in the hole channel AR lifetime. Furthermore, the decrease of the
hole channel AR lifetime was caused by an increase in the average
Coulomb coupling matrix elements ($V_{ijck}$) of the hole channel
and not by any substantial changes in the density of final states
(Table SX). Thus, it appears that the hole channel is more sensitive
to stress induced structural changes. And this suggests it is worthwhile
to perform more comprehensive studies on the impact of strain on AR
lifetimes, as strain may be playing a role in other nonmonotonic dependencies
of AR in core/shell nanomaterials.\citep{Pelton2017} That being said,
we do note that this is a rather small change of the AR lifetime and
experimental confirmation of this decrease in the AR lifetime upon
ZnS shell growth on QDs would likely be impeded by inhomogeneous broadening
and alloying of the core/shell interface. A more elaborate discussion
of the decrease of the AR lifetime upon the addition of a ZnS shell
and its relation to strain is given in the Supporting Information.

The goals of this study were to elucidate how biexciton Auger recombination
in colloidal core/shell QDs can be accurately modeled and efficiently
computed, and to uncover some of the underlying physics of excitons
and biexcitons in core/shell QDs by testing different approximations.
In order to achieve these goals, we developed a stochastic computational
scheme for calculating the nonradiative decay rate of biexcitonic
states. This efficient, stochastic method for calculating Auger recombination
lifetimes presented in this Letter is general and can be used for
any confined nanomaterial. We also utilized this efficient method
for calculating quantitatively accurate biexciton Auger recombination
lifetimes within an interacting (exciton--based) formalism to elucidate
the different impact of growing quasi--type--II (CdS) and type--I
(ZnS) shells on QD cores (CdSe). Specifically, we showed that the
Auger recombination lifetime monotonically increases as the number
of quasi--type--II shell monolayers increases whereas the Auger
recombination lifetime is mainly unchanged upon the addition of type--I
shells.
\begin{acknowledgments}
E.R. acknowledges support from the Department of Energy, Photonics
at Thermodynamic Limits Energy Frontier Research Center, under Grant
No. DE-SC0019140. We also acknowledge the University of California
Lab Fee Research Program (Grant LFR--17--477237) and the National
Energy Research Scientific Computing Center (NERSC), a U.S. Department
of Energy Office of Science User Facility operated under Contract
No. DE--AC02--05CH11231.
\end{acknowledgments}

\end{document}